\documentclass[twocolumn,prd,showpacs,amsmath,floatfix]{revtex4}

\usepackage{graphicx}
\usepackage{dcolumn}
\usepackage{bm}

\begin{document}

\title{
Air-shower simulations with and without thinning: artificial fluctuations
and their suppression }

\author{
D.S.~Gorbunov$^1$,
G.I.~Rubtsov$^{1,2}$
and S.V.~Troitsky$^1$}

\affiliation{$^1$Institute  for Nuclear Research of the Russian
  Academy of Sciences,
  Moscow 117312, Russia \\
$^2$Faculty of Physics, M.V.~Lomonosov Moscow State
  University,
  Moscow 119992, Russia
}

\date{March 16, 2007}

\begin{abstract}
The most common way to simplify extensive Monte-Carlo simulations of air
showers is to use the thinning approximation. We study its effect on
the physical parameters reconstructed from simulated showers. To this end,
we have created a library of showers simulated without thinning with
energies from $10^{17}$~eV to $10^{18}$~eV, various zenith angles and
primaries. This library is publicly available. Physically
interesting applications of the showers simulated without thinning are
discussed. Observables reconstructed from these showers are compared to
those obtained with the thinning approximation. The amount of artificial
fluctuations introduced by thinning is estimated. A simple method,
multisampling, is suggested which results in a controllable suppression of
artificial fluctuations and at the same time requires less demanding
computational resources as compared to the usual thinning.
\end{abstract}


\maketitle

\section{Introduction}
\label{sec:intro}

Experimental information about cosmic particles at very high energies
is obtained through the study of atmospheric showers induced by these
particles and is hence indirect. A necessary ingredient of these studies is
therefore a good understanding of a shower initiated by a primary
particle with given parameters. Since the shower development is a
complicated random process, the Monte-Carlo simulations are often used to
model atmospheric showers\footnote{
A completely different approach~\cite{Dedenko1968},
alternative to the full Monte-Carlo simulations, is to combine partial
Monte-Carlo with analytical solutions of cascade equations and
pre-simulated subshower libraries in the framework of {\it hybrid
codes}.}. Physical parameters are then reconstructed from the
simulations and compared to real data.

At very high energies, however, the number of particles in a shower is so
large that the simulations start to require unrealistic computer
resources. Among several ways to simplify the problem and to reduce the
computational time, the thinning approximation~\cite{Hillas:1997tf}
is currently the most popular one. Its key idea is to
track only a representative set of particles; while very efficient in
calculations and providing correct values of observables on average, this
method introduces artificial fluctuations because the number of tracked
particles is reduced by several orders of magnitude. These artificial
fluctuations mix with natural ones and therefore reduce the precision of
determination of physical parameters.

The standard approach to account for natural fluctuations in the
air-shower simulations is to fix all shower parameters and to simulate
sufficient number of artificial showers. Technically, these showers
differ by initial random seed numbers. All interactions in a
simulated shower are fixed by these numbers for a given thinning
level.  Random variations of these numbers result in a plethora of
possible interaction patterns which end up in a distribution of an
observable quantity of interest calculated for the showers with
exactly the same initial physical parameters. This distribution thus
intends to represent intrinsic fluctuations in the shower development.
Both the central value and the width of this distribution are important
for physical applications.

In practice, however, the width of the distribution arises from two
sources: physical fluctuations and artificial fluctuations
introduced by thinning. To obtain the physical width alone,
one should in principle perform simulations without thinning. This is
hardly possible for the highest energies at the current level
of computational techniques since one often needs to
simulate thousands of events for a typical study.

The aim of the present work is to estimate the relative size of these
artificial fluctuations (for the first time it is done by direct
comparison of showers simulated with and without thinning) and to develop
an efficient resource-saving method to suppress them in realistic
calculations.

In Sec.~\ref{sec:thinning}, we start with a description
(Sec.~\ref{sec:standard-thin}) of the standard thinning algorithm and
explain why its use introduces additional fluctuations. Then, we
briefly recall, in Sec.~\ref{sec:suppress-fluct}, conventional
approaches to avoid or suppress these
fluctuations. Sec.~\ref{sec:library} describes the library of showers
simulated without thinning for this study. This library is publicly
available.  Sec.~\ref{sec:fluct-size} is devoted to a quantitative
study of the artificial fluctuations. A new method, multisampling,
which allows one to suppress efficiently these unphysical fluctuations
without invoking extensive computer resources, is suggested and
discussed in Sec.~\ref{sec:msampl}.  Sec.~\ref{sec:concl} contains the
discussion of the method and our conclusions.


\section{Thinning approximation and beyond}
\label{sec:thinning}

\subsection{Standard thinning}
\label{sec:standard-thin}
The number of particles in an extended air shower (EAS), and hence
the CPU time and disk space required for its full simulation, scales
with the energy of the primary particle. At energies
in excess of $10^{17}$~eV, the number of particles of kinetic energy above
100~MeV at the ground level exceeds $10^8$ and the time required to
simulate such a shower at a computer with a few-GHz CPU is of order of
several days. A typical vertical shower induced by a hadron of
$10^{18}$~eV requires about 100~Gb of disk space and a month of CPU time.
Modelling individual showers with incident energies of about $10^{20}$~eV
is at the limit of realistic capabilities of modern computers; meanwhile
one needs thousands of simulated showers for comparison with
experimental data.

As a result of a full simulation of a shower, one obtains the list of all
particles at the ground level. This information is redundant for many
practical purposes. Real ground-based experiments detect only a small
fraction of these particles, so for calculating average particle densities
one does not need to know precise coordinates and energies of all
particles. In the thinning
approximation~\cite{Hillas:1997tf,Nagano:1999xk}, groups of particles are
replaced by effective representative particles with weights.

Let us briefly recall how the thinning approximation works (see e.g.\
Ref.~\cite{Kobal} for a detailed discussion). Denote the primary energy by
$E_0$ and introduce a parameter $\epsilon $ called the thinning
level. For each subsequent interaction, consider the energies $E_j$ of the
secondary particles created in this interaction. If the condition
\begin{equation}\label{thin0}
\sum E_j < \epsilon E_0
\end{equation}
is satisfied, then the method prescribes to keep one of the secondary
particles and to discard the others. The probability to keep the $i$th
particle $p_i$ is proportional to its energy,
$$
p_i = E_i / \sum_j E_j.
$$
To the selected particle, the weight $w_i=w_0/p_i$ is assigned, where
$w_0$ is the weight of the initial particle of this interaction ($w_0=1$
for the particle which initiated the shower).

If the condition~(\ref{thin0}) is not satisfied, then the so-called
statistical thinning operates: among the secondary particles, a subsample
of ones with energies $E_{j'}<\epsilon E_0$ is considered and
(one or more) effective particles are selected with probabilities
$$
p_{i^\prime} = \frac{E_{i^\prime}}{\min(\epsilon E_0,\sum_{j^\prime} E_{j^\prime})}~.
$$
These particles, to which the
weights $w_{i'}=w_0/p_{i'}$ are assigned,
are kept for further simulations together with original particles which
have had energies $E_{j'}>\epsilon E_0$.

For useful values of $\epsilon $, the number of particles tracked is
reduced by a factor of $10^3$ -- $10^6$. For a random process, this change
in the number of particles (and consequently, in the number of
interactions) results in the increase of fluctuations compared to the fully
simulated process.
This means that a part of
fluctuations in the development of a shower simulated with thinning is
artificial, that is it is present neither in
the full shower simulated with $\epsilon =0$ nor in a real EAS. For a
number of applications, these fluctuations are undesirable and should be
suppressed or at least brought under control.

\subsection{Standard methods to suppress fluctuations}
\label{sec:suppress-fluct}
In the framework of the thinning method, the fluctuations are
effectively suppressed by introducing the upper limit on the weight
factor $w_i$~\cite{Kobal}. The number of ``real'' particles
tracked is thus enlarged. Maximal weights for hadrons and for
electromagnetic particles may be assigned in different ways. For a given
problem, the optimal values of the maximal weights may be found in order to
minimise the ratio of the size of artificial fluctuations to the
computational time. In what follows, when we refer to the thinning with
weights limitation, we will use the maximal weights optimised in
Ref.~\cite{Kobal} for the calculation of the particle density.

The optimal values
of parameters of thinning procedure may depend on the interaction
models adopted in simulations for a given problem. In principle, the
weights should be optimized for each combination of the models (which are
updated every few years) and for each particular task (different
observables, primaries, energies, etc.). However, this optimization
requires a dedicated time consuming study in each case. We suggest
another approach to the problem in Sec.~\ref{sec:msampl}.

\subsection{A library of showers simulated without thinning}
\label{sec:library}
We have performed simulations of air showers without thinning by making use
of the CORSIKA simulation code~\cite{Heck:1998vt}. For different showers,
we have used QGSJET~01C~\cite{Kalmykov:1997te} and
QGSJET~II-03~\cite{Ostapchenko:2004ss} as high-energy and
GHEISHA~2002d~\cite{Fesefeldt:1985yw}  as low-energy hadronic interaction
models.
Currently, the library contains about 40 showers
induced by primary protons,
gamma-rays and iron nuclei
with
energies between $10^{17}$~eV and $10^{18}$~eV and zenith angles between
$0^\circ$ and $45^\circ$.
The showers have been simulated for the observational conditions
(atmospheric depth and geomagnetic field) of either
AGASA~\cite{Chiba:1991nf} or the Telescope Array~\cite{Kakimoto:2003ds}
experiments. The shower library~\cite{Rubtsov:GZK40} is publicly available
at {\tt http://livni.inr.ac.ru}. Detailed information about input
parameters used for the simulation of each shower is available from the
library website together with full output files. The access to the data
files is provided freely upon request. For users not familiar with
the CORSIKA output format, a "Datafile reading programming manual" is
given, containing a working example in C++. Free access to the
computational resources of the server is provided to avoid lengthy copying
of the output files (some of which exceed 100~Gb in size). An access
request form along with conditions of the usage of the library are
available from the library website.

Given the amount of computing resources required for simulation, each
shower simulated without thinning is valuable. We hope that the open
library would be useful in studies of various physical problems, notably facing
the improved precision of modern experiments which often exceeds the
precision of simulations.
The library is being
continuously extended; we plan to supplement it with showers of
higher energies in the future.



\section{Size of artificial fluctuations due to thinning}
\label{sec:fluct-size}
\subsection{Shower-by-shower comparison}
\label{sec:fluct-size-single}

\begin{figure}
\centerline{
\includegraphics[width=0.95 \columnwidth]{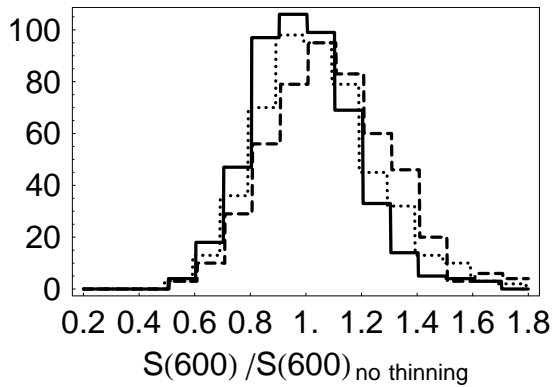}
}
\caption{
Distribution of $S(600)/S(600)_{\rm no~thinning}$, where $S(600)$
is
reconstructed from 500
showers simulated with $\epsilon =10^{-4}$ and the same random seed as the
corresponding $\epsilon =0$ shower, for
three different random seeds (three histograms). The three showers are all
vertical, induced by $10^{18}$~eV protons at the AGASA location.
\label{fig:distrib}
}
\end{figure}

\begin{figure}
\centerline{
\includegraphics[width=0.95 \columnwidth]{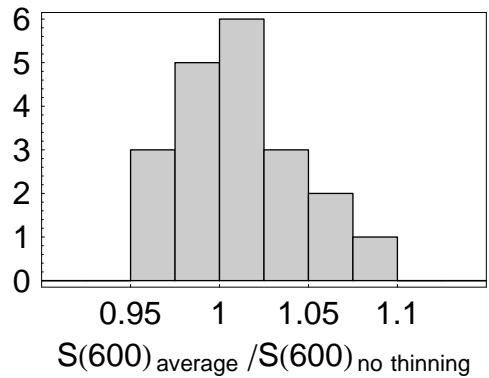}
}
\caption{
Distribution of $S(600)_{\rm average}/S(600)_{\rm
no~thinning}$, where $S(600)_{\rm average}$ is the average of
the reconstructed $S(600)$ over a sample of 500 showers simulated with
$\epsilon =10^{-4}$ and the same random seed as the corresponding $\epsilon
=0$ shower, for 20 different random
seeds. The 20 showers are all vertical, induced by $10^{17}$~eV protons at
the Telescope-Array location.
\label{fig:distrib20}
}
\end{figure}

\begin{figure}
\centerline{
\includegraphics[width=0.95 \columnwidth]{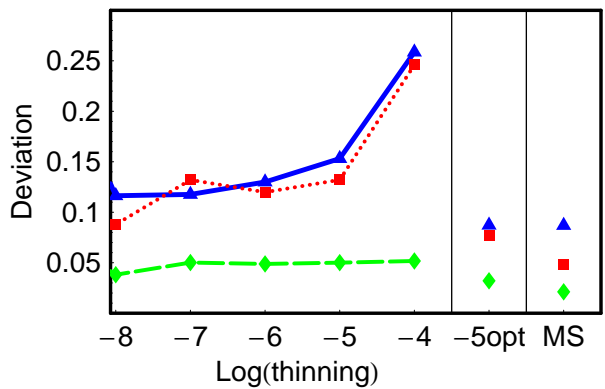}
}
\caption{
RMS deviations from unity of $S(600)/S(600)_{\rm
no~thinning}$ (blue triangles and thick blue line), $\rho _\mu (1000)/\rho
_\mu (1000)_{\rm no~thinning}$ (red boxes and red dotted line), $X _{\rm
max}/X_{\rm max,~no~thinning}$ (green diamonds and green dashed line),
where quantities with subscript ``no thinning'' are reconstructed from a
$10^{18}$~eV proton shower with zenith angle 45$^\circ$, simulated without
thinning for the AGASA observational conditions, while the rest of
quantities are reconstructed from large samples of showers simulated with
various thinning levels for the same input parameters and the same initial
random seed. ``$-5$ opt'' denotes $\epsilon =10^{-5}$ with weights
limitation; ``MS'' denotes multisampling ($20\times 10^{-4}$) discussed in
Sec.~\ref{sec:msampl}.
\label{fig:fluct1a}
}
\end{figure}

With a library of showers simulated without thinning, the
comparison of the observables reconstructed from showers with and without
thinning is possible. This allows one to estimate the effect of the
approximation. To do that, for each shower without thinning ($\epsilon=0$)
we have simulated a number of showers with different thinning levels
($\epsilon \ne 0$). All initial parameters (including the random seed
numbers) were kept the same as in the $\epsilon=0$ simulation, which
enabled us to reproduce exactly the same first interaction in the entire
set of showers. Three important observables --- the signal density at
600~m from the shower axis $S(600)$, the muon density at 1000~m from the
axis $\rho _\mu (1000)$, and the depth of the maximal shower development
$X_{\rm max}$ --- were reconstructed for each of the showers following the
data-processing operation adopted by the AGASA experiment\footnote{For the
Telescope Array, we have used the same procedure as for the AGASA
experiment with straightforward modifications taking into account the
thickness of scintillator detectors.}.  The detector response was
calculated with the help of GEANT simulations in Ref.~\cite{Sakaki}.
$S(600)$ and $\rho_\mu (1000)$ were obtained by fitting the corresponding
density at the ground level with empirical
formulae~\cite{Yoshida:1994jf,Hayashida:1995tu}. For fitting purpose the
density was binned into 50m-width rings centered at the shower axis.
$X_{\rm max}$ was obtained by fitting the longitudinal shower profile with
the empirical Gaisser-Hillas curve~\cite{GaisserHillas} (incorporated into
CORSIKA). This procedure was repeated for all showers in the Livni library
with the results similar to those shown in
Figs.~\ref{fig:distrib}--\ref{fig:fluct1a}.

Figure~\ref{fig:distrib}
shows the distribution of the reconstructed $S(600)$ for showers
with thinning simulated with the same initial random seed (and thus the
same first interaction) as three representative $\epsilon =0$ Livni
showers. Though quite wide for $\epsilon =10^{-4}$ thinning, the
distributions of
$S(600)/S(600)_{\rm no~thinning}$
are well centered at unity.

The distribution of the mean values of
$S(600)/S(600)_{\rm no~thinning}$
for the ensembles of the thinned showers is presented in
Fig.~\ref{fig:distrib20}
for a uniform sample of twenty different $\epsilon =0$ showers. For each
of them, 500 showers with $\epsilon =10^{-4}$ were simulated with the same
first interaction as the corresponding $\epsilon =0$ shower. The
values of the observable averaged over 500 thinned showers approximate the
``exact''
$S(600)_{\rm no~thinning}$
with the accuracy of about $3\%$, which is consistent with the level of
statistical fluctuations, $1/\sqrt{500}\sim 4\%$. We have found the same
distributions for other observables considered, $\rho _\mu (1000)$ and
$X_{\rm max}$. The important
conclusion is that for the first time, {\em the usual assumption that
thinning does not introduce systematic errors in the reconstructed
observables has been checked by explicit comparison of $\epsilon=0$ shower
and averaged $\epsilon \ne 0$ showers}, at least for energies up to
$10^{18}$~eV, observables $S(600)$, $\rho _\mu (1000)$ and $X_{\rm max}$,
and proton, photon and iron primaries.

The spread of observables reconstructed from thinned showers depends on the
thinning level $\epsilon $. It is not the width of the
distribution but the average deviation of the observables from those of an
$\epsilon =0$ shower which is the most interesting for practical purposes.
This quantity is plotted in Fig.~\ref{fig:fluct1a}
for a typical shower from the Livni library.

We note in passing that, technically, to study
the spread at a given $\epsilon $ with CORSIKA,
one has to simulate showers with
slightly different thinning levels (otherwise they all would be
absolutely identical, given a fixed random seed). For instance, to
obtain the points corresponding to $\epsilon=10^{-5}$ in
Fig.~\ref{fig:fluct1a}, we have simulated 500 showers with different
thinning levels in the interval $0.99\cdot 10^{-5}<\epsilon <1.01\cdot
10^{-5}$.

\subsection{Distributions of showers}
\label{sec:fluct-size-ensemble}
In most cases one is not interested in details of a particular
realization of a shower; it is the ensemble of simulated showers with
fixed initial parameters but varied random seeds which is compared to
the real data. The study of Sec.~\ref{sec:fluct-size-single} does not
help seriously to estimate the effect of thinning on these
distributions of parameters because the size of fluctuations seen,
e.g., in Fig.~\ref{fig:fluct1a} is determined by a combination of
artificial fluctuations and a part of real ones: the random seed
together with initial conditions fixes the first interaction, but
different thinning levels introduce variations in other interactions
and effectively change the simulation of the entire shower development.

To estimate the effect of thinning on the distribution of observables, we
have simulated samples of showers with fixed initial conditions but
different random seeds for various thinning levels, including
$\epsilon=0$. We have considered samples of $E=10^{17}$~eV vertical
proton-induced showers consisting of 20 showers with $\epsilon=0$, 100
showers with $\epsilon=10^{-5}$, 100 showers with $\epsilon=10^{-5}$ and
weight limitation, 100 showers with $\epsilon=10^{-4}$ and 100 showers
with $\epsilon=10^{-4}$ and weight limitation.

\begin{figure}
\centerline{
\includegraphics[width=0.95 \columnwidth]{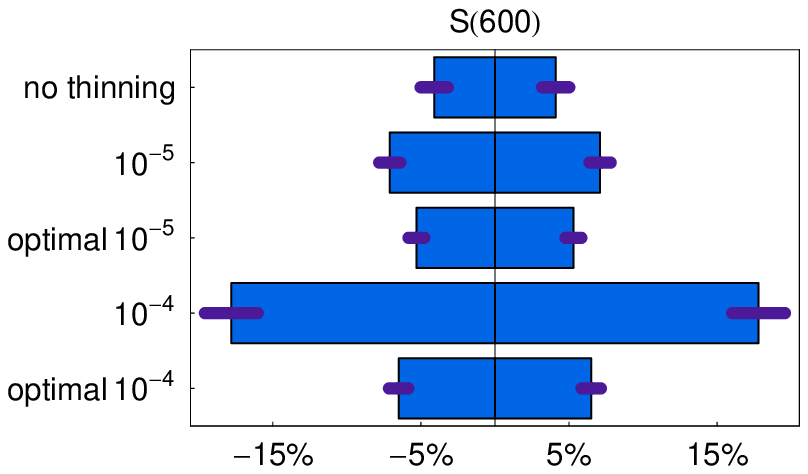}
}
\centerline{
\includegraphics[width=0.95 \columnwidth]{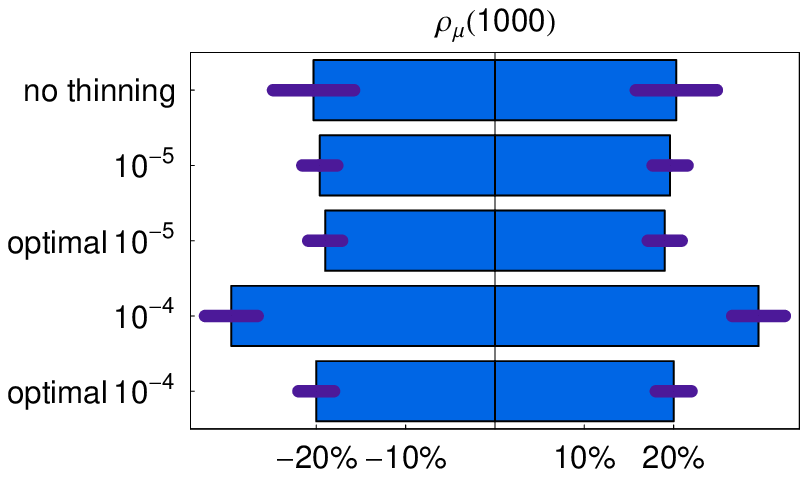}
}
\caption{
\label{fig:TA1e17s}
Width of the $S(600)$ distribution (upper panel) and $\rho _\mu
(1000)$ distribution (lower panel) for $10^{17}$~eV vertical proton
showers simulated with and without thinning for the Telescope Array
observational conditions. Statistical error bars are due to a limited
number of simulated showers. }
\end{figure}

The simulations have been performed using QGSJET~II and GHEISHA as hadronic
interaction models, for the observational conditions of the Telescope Array
experiment. The distributions of $S(600)$, $\rho _\mu (1000)$ and $X_{\rm
max}$ have been reconstructed with statistical fluctuations
(originated from the limited number $n$ of showers in the samples) of
about $1/\sqrt{n}$, that is, about 23\% for $\epsilon=0$ showers and about
10\% for the other samples. Figure \ref{fig:TA1e17s}
illustrates the widths of the
distributions obtained at different $\epsilon $.
Artificial fluctuations in $S(600)$ and $\rho _\mu (1000)$ caused by
thinning are clearly seen by comparing $\epsilon=10^{-4}$ case with others
(for $X_{\rm max}$ the artificial fluctuations are quite small).
We note that, for a given $\epsilon $, the fluctuations should be larger
at high energy since the multiplicity of hadronic interactions grows with
energy and thinning starts to operate earlier in the shower affecting the
first few interactions which determine the fluctuations. For many
practical purposes, these artificial fluctuations should be efficiently
suppressed.


\section{Multisampling: an economical method to suppress artificial
fluctuations}
\label{sec:msampl}

From the results of the previous section, we conclude that the use of
thinning is well motivated when one is interested in the reconstruction of
the central values of fluctuating observables (the most important
application is e.g.\ to establish a relation between, say, $S(600)$ and
energy for a given experimental setup). On the other hand, thinning may
limit the precision of composition studies, where the observed value of
some quantity is compared to the simulated distributions of the same
quantity for different primaries, and the width of these distributions is
of crucial importance (see e.g.\ the proton--iron comparison in examples
of Ref.~\cite{our-composition}).

\begin{figure}
\centerline{
\includegraphics[width=0.95 \columnwidth]{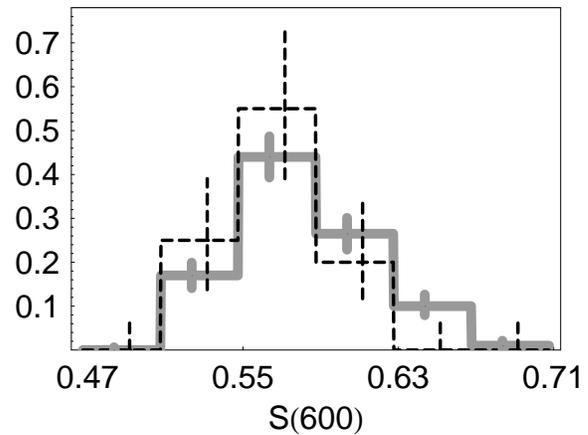}
}
\caption{
\label{fig:s600distrib2}
Normalized distributions of the reconstructed $S(600)$ (in VEM/m$^2$):
dashed line, 20 showers without thinning; solid line, 500 showers with
$(20\times 10^{-4})$ multisampling. All showers are vertical, induced
by $10^{17}$~eV primary protons simulated for the conditions of the
Telescope Array. Vertical error bars are due to limited statistics. }
\end{figure}

As it has been pointed out above, the effect of physical fluctuations on
the distribution of an observable quantity should be, in principle,
estimated by simulating a set of showers with the same physical
parameters, with different random seeds and without thinning. To obtain a
good approximation to this distribution,
we make use of the results of
Sec.~\ref{sec:fluct-size-single} (see, in particular,
Figs.~\ref{fig:distrib} and \ref{fig:distrib20}).
The average of an observable over
a sample of thinned showers with a fixed initial random seed approximates
the value of the same observable for an $\epsilon =0$ shower with the same
random seed with a good accuracy. The distribution of observables for
$\epsilon =0$ showers with different random seeds is then approximated by
a distribution of these approximated observables calculated for samples
with random seeds varying from one sample to another but fixed inside a
sample. A practical way to do this is as follows:
\begin{itemize}
 \item
instead of a single shower with $\epsilon=0$, simulate $N$ showers
with some $\epsilon=\epsilon_0\neq 0$ and fixed random seed;
\item
reconstruct the observable for each of $N$ showers, average over these $N$
realizations and keep this average value which approximates the result for
a single shower without thinning;
\item
repeat the procedure $M$ times for different random seeds to mimic a
simulation of $M$ showers without thinning and to obtain the required
distribution of the observable.
\end{itemize}
We will refer to this procedure as to {\em multisampling} $(N\times
\epsilon_0)$.  Even for relatively large $\epsilon $, averaging over a
sufficiently large number of showers ($N$) gives a good approximation
to an $\epsilon_0=0$ value of the observable; the larger $N$, the better
the approximation. The required value of $N$ may be estimated as
follows. Consider the distribution of an observable reconstructed from
showers simulated with the thinning level close to $\epsilon_0$ for a
given initial random seed. Assume that the distribution is Gaussian
with the width $\sigma $ (though the qualitative conclusions do not
depend on the exact form of the distribution, we note that, in practice,
it is indeed very close to Gaussian~\cite{ST:GZK40}); then one needs
$N$ measurements to know the mean value with the precision $\sim\sigma
/\sqrt{N}$. Numerical results for the Livni showers demonstrate that
$(N\times \epsilon_0)$ multisampling for $N\sim 15 \dots 20$ and
$\epsilon_0\sim 10^{-4}$ results in the precision of $3\div 4\%$ in the
reconstruction of $S(600)$, $\rho _\mu (1000)$ and $X_{\rm max}$ of
the original $\epsilon=0$ showers. The distributions
of parameters reconstructed from showers without thinning are
consistent (within statistical errors) with those extracted by making
use of multisampling. The distributions of $S(600)$ are presented in
Fig.~\ref{fig:s600distrib2}.

\begin{figure}
\centerline{
\includegraphics[width=0.95 \columnwidth]{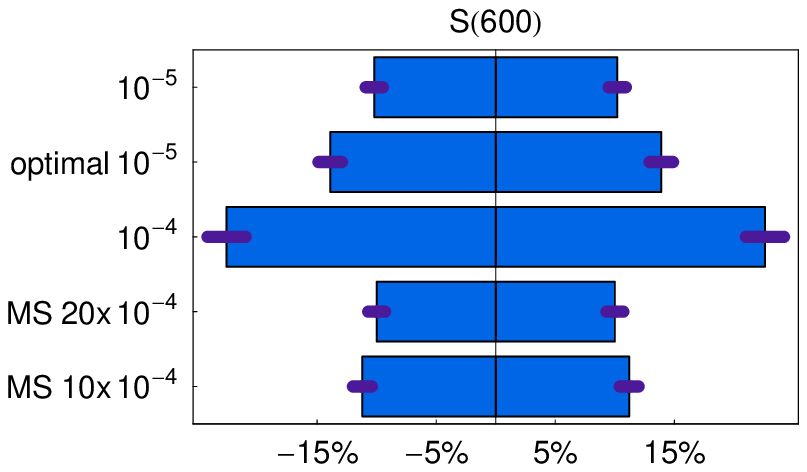}
}
\centerline{
\includegraphics[width=0.95 \columnwidth]{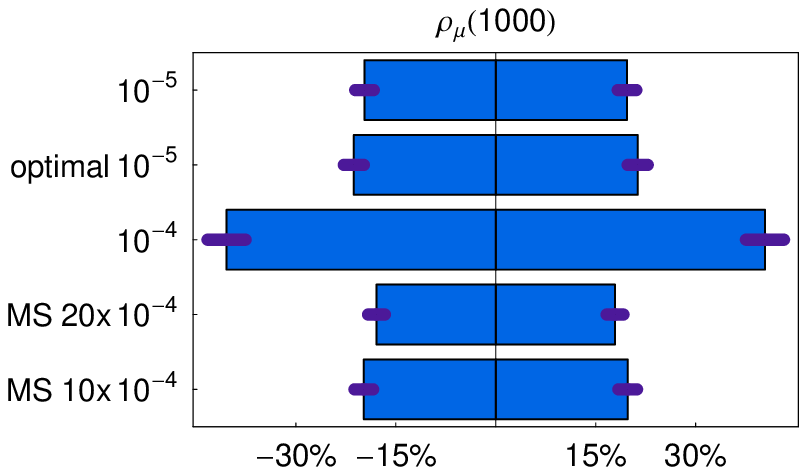}
}
\caption{
\label{fig:TA5e19s}
Width of the $S(600)$ distribution (upper panel)
and $\rho _\mu (1000)$ distribution (lower panel) for 200 showers initiated
by $5\cdot 10^{19}$~eV vertical protons simulated with thinning and with
multisampling for the Telescope Array observational
conditions. Statistical error bars are due to a limited number of
simulated showers. The choice of maximal weights suggested in
Ref.~\cite{Kobal} may not be optimal for this study (see the text).}
\end{figure}

In
Fig.~\ref{fig:TA5e19s}
we present the widths
of the distributions obtained with the usual thinning and with
multisampling for
$E=5\cdot 10^{19}$~eV vertical proton-induced showers; the limited
statistics (we used $n=200$ showers) implies the statistical uncertainty
of about $1/\sqrt{n}\sim 7\%$. The gain in precision is clearly seen; for
the case of $5\cdot 10^{19}$~eV the multisampled distribution (which is
expected to mimic the $\epsilon=0$ distribution with a good accuracy)
allows us to estimate the size of purely artificial fluctuations caused by
thinning. For instance, for $\epsilon =10^{-5}$ with weights
limitations, these fluctuations remain at the level of $\gtrsim 10\%$ for
$S(600)$ and of $\gtrsim 12\%$ for $\rho _\mu (1000)$. Let us note in
passing that, for this particular simulation ($5\cdot 10^{19}$~eV vertical
protons at the Telescope Array location) and for our choice of hadronic
models (QGSJET~II and GHEISHA), the choice of maximal weights suggested in
Ref.~\cite{Kobal} may not be optimal.

Let us compare now the computer resources needed for calculations with the
standard thinning (with and without weights limitations) and with
multisampling.

\begin{figure}
\centerline{
\includegraphics[width=0.6 \columnwidth]{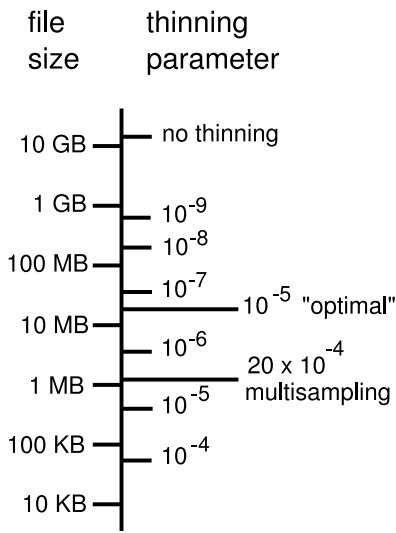}
}
\caption{
\label{fig:disk-versus-T}
The size of the CORSIKA output file for different thinning levels and
procedures (a $10^{18}$~eV proton shower with zenith angle of
$45^\circ$, AGASA observational conditions).}
\end{figure}

The disk space scales as the number of simulated particles;
Fig.~\ref{fig:disk-versus-T}
illustrates this fact. We see that the
multisampling ($20 \times 10^{-4}$) saves the disk space compared to
$\epsilon=10^{-5}$ with weights limitation, giving at the same time gain in the
precision of simulations.

The CPU time is very sensitive to the choice of the hadronic
interaction model: since thinning starts to work when the number of
particles is large enough, the first few interactions are simulated in
full even for relatively large $\epsilon $. If the high-energy model
is slow, then the effect of multisampling on the computational time is
not so pronounced. By variations of the hadronic interaction models,
we have estimated the average time consumed by QGSJET~II, SYBILL, FLUKA and
GHEISHA for simulations of showers at energies $10^{17}$~eV and
$5\cdot10^{19}$~eV. For $5\cdot10^{19}$~eV vertical proton showers, ($20
\times 10^{-4}$) multisampling is about 5 times faster than $10^{-5}$
thinning with weights limitation for SYBILL while for (very slow)
QGSJET~II, both take roughly the same time.
A way to change the multisampling procedure in order to
gain in the CPU time for any hadronic model is discussed below in
Sec.~\ref{sec:concl}.


\section{Discussion and conclusions}
\label{sec:concl}
A library of atmospheric showers has been simulated without the thinning
approximation. The showers have been used for a quantitative
direct study of the effect of thinning on the reconstruction of signal
($S$) and muon $(\rho _\mu $) densities at the ground level as well as
on the depth $X_{\rm max}$ of the maximal shower development. We
demonstrate that thinning {\it does not introduce systematic shifts }
into these observables, as was conjectured but never explicitly
checked. We estimate the size of artificial fluctuations which appear
due to the reduction of the number of particles in the framework of
the thinning approximation; these unphysical fluctuations may affect
the precision, e.g., of the composition studies. For instance, at the
energies of $5\cdot 10^{19}$~eV for vertical proton primaries,
artificial fluctuations are about 10\% in the signal density at 600~m
and about 12\% in the muon density at 1000~m for $\epsilon=10^{-5}$
thinning with weight limitations. An effective method to suppress
these artificial fluctuations, multisampling, is suggested and
studied. The method does not invoke any changes in simulation codes;
only parameters of, say, the CORSIKA input are affected.  Compared to
the $10^{-5}$ thinning with weights limitations, it gives a similar
precision but allows one to gain an order-of-magnitude decrease in the
required disk space. Gain in the CPU time depends on the speed of the
high-energy interaction model: it is of order $5\div 10$ for fast ones
(SYBILL) and of order 1 for slow ones (QGSJET~II).

A way to change the multisampling procedure in order to further
improve the gain in the CPU time is to simulate the high-energy part
of a shower once for each initial random seed while having the
low-energy part multisampled. The multisampling procedure described
above is a particular case of such improved procedure with a high-energy
part restricted to the first interaction only. We would expect the
modification to make it possible to conserve the physical fluctuations
in the second and several following interactions and will allow for an
order-of-magnitude improvement in the computational time for any hadronic
model. However, it would require (simple) changes in the simulation codes
thus loosing an important advantage of the multisampling discussed above:
to implement the latter, one operates with the standard simulation code
(e.g., CORSIKA) without any modifications. This minimal change
is to add the option to start simulations from a {\em
predefined set } of the primary particles.

We are indebted to T.I.~Rashba and V.A.~Rubakov
for helpful discussions.
This work was supported in part by the INTAS grant 03-51-5112, by
the Russian Foundation of Basic Research grants 07-02-00820,
05-02-17363 (DG and
GR), by the grants of the President of the Russian Federation
NS-7293.2006.2 (government contract 02.445.11.7370; DG, GR and ST) and
MK-2974.2006.2 (DG) and by the Russian Science Support
Foundation (ST). Numerical part of the work was performed at the computer
cluster of the Theory Division of INR RAS. Our library of showers without
thinning is publicly available at {\tt http://livni.inr.ac.ru}.


\end{document}